\documentclass[
    ,final            
  ]
  {aipproc}
\layoutstyle{6x9}

\usepackage{amssymb}

\newcommand{\apj}{ApJ}
\newcommand{\apjs}{ApJS}
\newcommand{\apjl}{ApJ}
\newcommand{\araa}{ARA\&A}
\newcommand{\aap}{A\& A}

\newcommand{\physrep}{Phys.~Rep.}

\begin{document}

\title{What can we learn from long term monitoring of X-ray bursters?}


\author{Andrew Cumming}{
  address={Physics Department, McGill University, 3600 rue University, Montreal QC, H3A 2T8, Canada}
}

\begin{abstract}
The last few years have seen the discovery of a number of new aspects of Type I X-ray bursts: the extremely energetic and long duration superbursts, intermediate duration bursts at low luminosities, mHz QPOs, and burst oscillations. These discoveries promise a new understanding of nuclear burning on accreting neutron stars, and offer a chance to use observations to probe neutron star properties.  I discuss what we can learn from future long term monitoring with MIRAX.
\end{abstract}

\maketitle

\section{Introduction}

The fate of matter accreted onto a neutron star is an old question \cite{Rosenbluth73, Hansen75}, but also an important one. A neutron star in a low mass X-ray binary (LMXB) likely accretes enough mass in its lifetime to replace the entire crust, affecting the long term spin, thermal, and magnetic field evolution of the star. Also, observations of nuclear processing of accreted material offer a chance to probe properties of the neutron star, such as interior thermal structure, magnetic field, and spin. Our understanding of nuclear burning on the surfaces of accreting neutron stars has improved significantly in the last few years. Observations of X-ray bursters with BeppoSAX and the Rossi X-ray Timing Explorer (RXTE) have revealed a host of new phenomena (see \cite{StrohBild03} for a review), including millisecond oscillations during Type I X-ray bursts, the long duration and energetic ``superbursts'', long duration X-ray bursts from faint sources, and mHz quasi-periodic oscillations in the persistent emission that are likely due to nuclear burning. In this article, I discuss two aspects which will be significantly impacted by long term monitoring with MIRAX.

\begin{figure}
\includegraphics[height=.35\textheight]{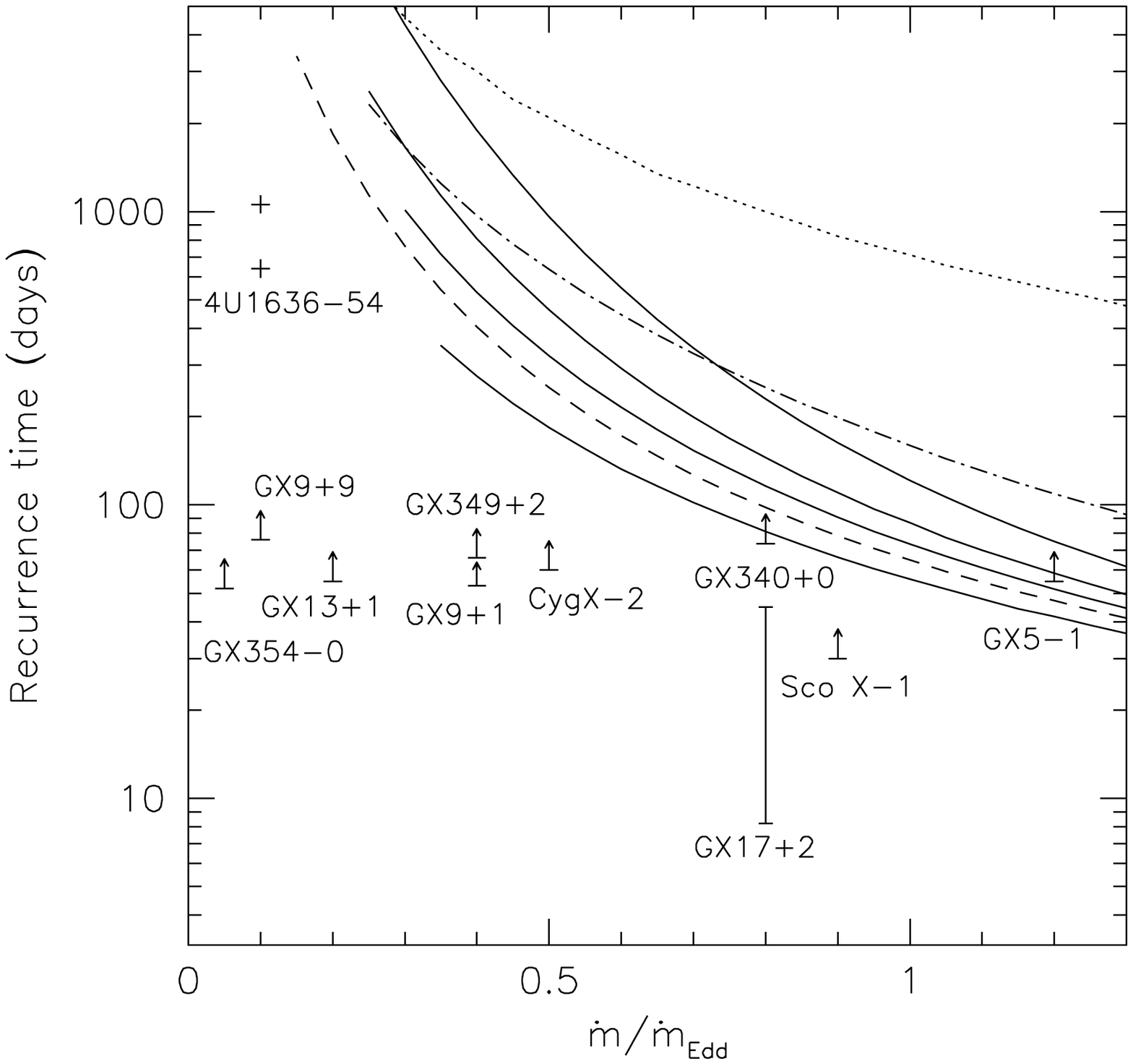}
\includegraphics[height=.35\textheight]{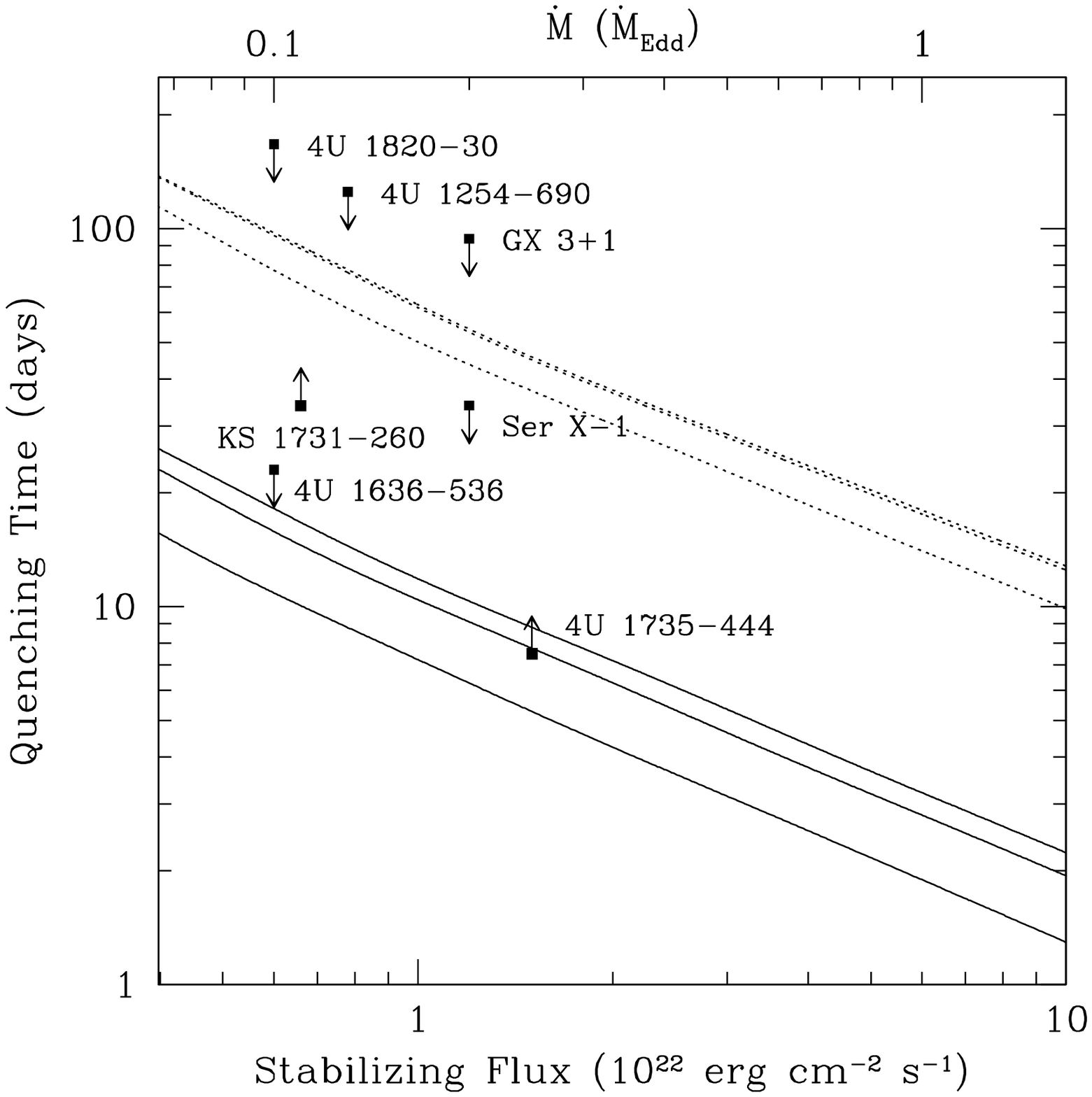}
\caption{Observations of superbursts that will be significantly improved by MIRAX. {\em Left panel:} Recurrence times compared to theoretical models (from \cite{Keek06}). 4U~1636-54 and GX~17+2 have shown multiple superbursts, allowing a measurement of recurrence time. For the  other sources, we show lower limits from BeppoSAX. {\em Right panel:} Quenching of normal Type I bursts following a superburst (from \cite{CM04}). We show the predicted quenching time for $y_b=10^{12}$ (solid lines) and $10^{13}\ {\rm g\ cm^{-2}}$ (dotted lines) as functions of the critical flux needed to stabilize H/He burning, and the accretion rate. For each value of $y_b$, the curves are for (bottom to top) energy releases of $1$, $2$ and $3\times 10^{17}\ {\rm erg\ g^{-1}}$. The data points are taken from Table 1 of \cite{Kuulk03}, with an updated value for 4U~1636-53 (Kuulkers, private communication).}
\end{figure}

\section{Probing neutron star interiors with superbursts}

One of the exciting developments in the last few years has been the discovery of superbursts. These are long duration (several hours), rare (recurrence times $\sim $ 1 year), and energetic ($10^{42}\ {\rm ergs}$) X-ray bursts  \cite{Kuulk03}, believed to be due to unstable burning of a thick layer of carbon. The basic idea is that the accreted hydrogen and helium burns within hours to days of arriving on the surface of the star, leaving behind a mixture of heavy elements and a small amount ($\sim 10$\% by mass) of carbon \cite{Schatz99, Schatz03}. This mixture accumulates until the carbon starts to burn, triggering the superburst \cite{CB01, SB02}.

Theoretical studies of superbursts initially focused on their potential as probes of nuclear physics. Hydrogen and helium burning proceeds by the rp-process \cite{WW81,Schatz98}, a series of proton captures and beta-decays involving heavy nuclei close to the proton drip line. This process naturally explains the $\approx 100\ {\rm s}$ extended tails observed in some X-ray bursts (e.g.~from the regular burster GS~1826-24 \cite{Galloway04,C04}). The properties and reaction rates of the unstable proton-rich nuclei involved in the rp-process are not well known experimentally \cite{Schatz98}. Because the heavy element composition sets the thermal conductivity of the accumulating carbon layer, Cumming \& Bildsten \cite{CB01} suggested that the superburst properties would directly probe the composition of the rp-process ashes.

Brown \cite{B04} and Cooper \& Narayan \cite{CN05} showed that in fact the ignition conditions are  much more sensitive to the thermal properties of the crust and core. This is exciting because it gives a new way to probe the neutron star interior, complementary to observations of transiently-accreting neutron stars in quiescence \cite{Rut02,Wijnands02,Yak04}, or cooling isolated neutron stars (see \cite{Yak04} for a recent review). They found that to match observed superburst recurrence times required inefficient neutrino cooling in the core, and a low thermal conductivity in the crust. 

We recently revisited this question in \cite{C05}, with two main improvements: (1) using models of superburst lightcurves to constrain the ignition depth, and (2) including neutrino emission from the crust from the formation of Cooper pairs \cite{Yak99}. Cumming \& Macbeth \cite{CM04} modeled superburst lightcurves, finding that the luminosity decays as a broken power law, with break time corresponding to the thermal time at the base of the layer. Fitting these model lightcurves to the observed superbursts constrains the ignition depth to be $\approx 10^{12}\ {\rm g\ cm^{-2}}$ \cite{C05}. The temperature required to ignite carbon at this depth is $\approx 6\times 10^8\ {\rm K}$. Without Cooper pair neutrino emission from the crust, we find that achieving this temperature requires a poor crust conductivity and the inefficient core neutrino emission (modified URCA or smaller), in agreement with previous work. However, surprisingly, including Cooper pair emission makes the crust too cold to achieve ignition at the inferred depth, even for very inefficient core neutrino emission and a low crust conductivity. {\em It is not possible to match the observed superburst ignition depth when Cooper pair cooling is included.}

The resolution to this puzzle is not yet clear, but will tell us something about the neutron star interior. The Cooper pair neutrino emissivity may be smaller than current calculations suggest, or there may be an extra heating source in the accumulating fuel layer that is not included in current models. An alternative explanation is that these stars are not neutron stars, but rather ``strange stars'' \cite{Alcock86}. Strange stars do not have an inner crust, and so naturally do not have Cooper pair neutrino emission. An evaluation of this scenario \cite{PC05} shows that ignition at the observed depth is possible for a wide range of parameters for the strange star core.

There are two aspects of superbursts where MIRAX will have a significant impact, providing new observational constraints to help answer these questions. The first is increasing the sample of known superbursts and measuring or improving limits on recurrence times. Figure 1 summarizes the current observational constraints on superburst recurrence times as a function of accretion rate (see \cite{Keek06}). The lower limits of $\approx 50$ days are set by the total exposure time accumulated by BeppoSAX, and will improve by an order of magnitude with MIRAX, allowing much better constraints on theoretical models. The second aspect is measuring quenching times of normal Type I bursts following a superburst. Type I bursts disappear for $\approx 1$ month following a superburst, as the heat flux from the carbon layer stabilizes the hydrogen and helium burning \cite{CB01}. Current observations are summarized in Figure 1, but are not very constraining. MIRAX will easily measure quenching timescales, providing an independent measurement of the thickness of the fuel layer and an additional constraint on the theoretical models \cite{CM04}.

\section{mHz QPOs and the transition to stable burning}

LMXBs exhibit a range of periodic and quasi-periodic phenomena, ranging in frequency from very low frequency (mHz) noise to kHz quasi-periodic oscillations (QPOs) (see \cite{vdk04} for a review). This variability has mostly been associated with orbiting material in the accretion flow close to the compact object. In the case of a neutron star accretor, an important question is whether any of these phenomena originate from or are associated with the neutron star surface. This is important for identifying the compact object as a neutron star or a black hole, as well as offering a probe of the neutron star surface layers. 

Revnivtsev et al.~\cite{Rev01} discovered a new class of mHz QPOs in three Atoll sources, 4U~1608-52, 4U~1636-53, and Aql X-1, which they proposed were from a special mode of nuclear burning on the neutron star surface rather than from the accretion flow. These mHz QPOs have frequencies in the range $7$--$9\ {\rm mHz}$ (timescales of $1.9$--$2.4$ minutes). In 4U~1608-52, a transient source  whose luminosity is observed to change by orders of magnitude, the mHz QPO was only present within a narrow range of luminosity, $L_X\approx 0.5$--$1.5\times 10^{37}\ {\rm erg\ s^{-1}}$. This is significant because in many X-ray bursters a transition in burning behavior occurs close to this luminosity, from frequent Type I X-ray bursting at low accretion rates to the disappearance of Type I X-ray bursts at high accretion rates (e.g.~\cite{Corn03}). This transition is expected theoretically because at high accretion rates the fuel burns at a higher temperature, reducing the temperature-sensitivity of helium burning and quenching the thin shell instability. However, an outstanding puzzle is that theory predicts a transition accretion rate close to the Eddington rate, almost an order of magnitude larger than observed \cite{B98}.

Paczynski \cite{Pac83} pointed out that near the transition from instability to stability, oscillations are expected because the eigenvalues of the system are complex (see also \cite{NH03}). Extending his analysis, we recently showed \cite{Heger06} that the mHz QPOs are indeed naturally explained as being due to marginally stable nuclear burning on the neutron star
surface. At the boundary between unstable and stable burning, the temperature
dependence of the nuclear heating rate and cooling rate almost
cancel. The result is an oscillatory mode of burning, with an
oscillation period close to the geometric mean of the thermal and
accretion timescales for the burning layer, or $\approx 100\ {\rm s}$, matching the observed periods.
Numerical simulations with the Kepler code confirm this simple analytic understanding.  Interestingly, the observed oscillation period depends sensitively on the surface gravity and the accreted hydrogen fraction, giving a new way to probe these parameters. We are currently working to understand in detail the range of luminosities for which mHz QPOs are observed, and what sets the Q value of the oscillation.

Improved understanding of mHz QPOs in particular, and the global burning behavior in general, from long term monitoring could tell us about the geometry of accretion onto the star. If the mHz QPOs are due to marginally stable nuclear burning, the local accretion rate onto the star must be close to the Eddington rate, even though the global accretion rate inferred from the X-ray luminosity is ten times lower. One  possibility is that the accreted material covers only part of the neutron star surface at luminosities $L_X\gtrsim 10^{37}\ {\rm erg\ s^{-1}}$. This would also explain a number of other puzzling features of Type I X-ray bursts. First, it would account for the disappearance of regular bursting at this luminosity. If some fuel  ``leaked out'' from the stably burning region, it could ignite and cause the occasional X-ray bursts seen at high luminosities. Stable H/He burning at global luminosities well below Eddington would also allow carbon production in large enough quantities to power superbursts \cite{Schatz03}. Burst oscillations occur preferentially at higher luminosities \cite{muno00}. It may be that incomplete spreading of fuel promotes the development of burning inhomogeneities. Finally, Atoll sources undergo a transition from the island to banana state of the color-color diagram close to this luminosity. Perhaps a change in accretion geometry affects the distribution of fuel.

\section{Conclusions}

In summary, long term monitoring of X-ray bursters with BeppoSAX and RXTE has revealed new phenomena that would be difficult to observe otherwise. I have concentrated on superbursts, which occur on long timescales, and mHz QPOs, which occur in a narrow luminosity range. Unfortunately, there is no space in this short article to discuss ``intermediate duration'' bursts, which are also sensitive to the neutron star interior properties (e.g.~\cite{intzand05}), low luminosity bursters (e.g.~\cite{Corn02}), burst oscillations, or how burst properties can be used to constrain the composition of the accreted material \cite{C04}. MIRAX will dramatically improve our understanding of all of these phenomena, which promises to tell us about (1) the physics of high density matter, (2) neutron star spin and magnetic field evolution, (3) the composition of the donor star and therefore evolution of low mass X-ray binaries, (4) nuclear physics at high temperatures and densities, and (5) the geometry of the accretion flow onto the star.

I would like to thank the conference organizers for their hospitality, and Duncan Galloway, Erik Kuulkers, and Jean in 't Zand for discussions. I acknowledge support from an NSERC Discovery Grant, Le Fonds Qu\'eb\'ecois de la Recherche sur la Nature et les Technologies, and the Canadian Institute for Advanced Research.

\end{document}